\documentclass[aps,prl,twocolumn,showpacs,superscriptaddress,groupedaddress]{revtex4}  % for review and submission
\usepackage{graphicx}  % needed for figures
\usepackage{dcolumn}   % needed for some tables
\usepackage{bm}        % for math
\usepackage{amssymb}   % for math
\usepackage{amsmath}   % for math
\begin{document}
\title{Dephasing in Mach-Zehnder interferometer by an Ohmic contact}

\author{Edvin G. Idrisov}
\affiliation{D\'epartement de Physique Th\'eorique, Universit\'e de Gen\`eve, CH-1211 Gen\`eve 4, Switzerland}

\author{Ivan P. Levkivskyi}
\affiliation{Theoretische Physik, ETH Zurich, CH-8093 Zurich, Switzerland}
\affiliation{Institute of Ecology and Evolution, University of Bern, CH-3012 Bern, Switzerland}
\affiliation{Department of Computational Biology, University of Lausanne, CH-1011 Lausanne, Switzerland}

\author{Eugene V. Sukhorukov}
\affiliation{D\'epartement de Physique Th\'eorique, Universit\'e de Gen\`eve, CH-1211 Gen\`eve 4, Switzerland}
\date{\today}

% The following information is for internal review, please remove them for submission
%\widetext
%\leftline{Version xx as of \today}
%\leftline{Primary authors: Joe E. Physics}
%\leftline{To be submitted to (PRL, PRD-RC, PRD, PLB; choose one.)}
%\leftline{Comment to {\tt d0-run2eb-nnn@fnal.gov} by xxx, yyy}
%\centerline{\em D\O\ INTERNAL DOCUMENT -- NOT FOR PUBLIC DISTRIBUTION}

% the following line is for submission, including submission to the arXiv!!
%\hspace{5.2in} \mbox{Fermilab-Pub-04/xxx-E}

%\title{Template for PRL/PRD Papers}
%\input author_list.tex       % D0 authors (remove the first 3 lines
                             % of this file prior to submission, they
                             % contain a time stamp for the authorlist)
                             % (includes institutions and visitors)
%\date{\today}

\begin{abstract}
We study dephasing in an electronic Mach-Zehnder (MZ) interferometer based on quantum Hall (QH) edge states by a micromiter-sized  Ohmic contact embedded in one of its arms. We find that at the filling factor $\nu=1$, as well as  in the case where an Ohmic contact is connected to an MZ interfeoremter by a quantum point contact (QPC) that transmits only one electron channel, the phase coherence may not be fully suppressed.  
Namely, if the voltage bias $\Delta \mu$ and the temperature $T$ are small compared to the charging energy of the Ohmic contact $E_C$, the free fermion picture is manifested, and the visibility saturates at its maximum value. At large biases, $\Delta \mu \gg E_C$, the visibility decays in a power-law manner. 
%At high temperatures, $T \gg E_C$, the visibility decays exponentially with temperature.   

   %We study dephasing in an electronic Mach-Zehnder interferometer (MZI) strongly coupled to an Ohmic contact. We find the visibility of Aharonov-Bohm (AB) oscillations as a function of the applied bias, $\Delta \mu$, and temperature, $T$. Low-bias, $\Delta \mu \to 0$, $T \neq 0$, and non-linear, $\Delta \mu \neq 0$, $T \to 0$, regimes are considered. In the low-bias regime at low temperatures, $T \ll E_C$,  where $E_C$ is the charging energy of the Ohmic contact, a free fermion picture is manifested , and the visibility does not depend on the temperature and charging energy. At high temperatures, $T \gg E_C$, the visibility decays exponentially with temperature. In the non-linear regime at small biases, $\Delta \mu \ll E_C$, the free fermion behavior is restored. At large biases, $\Delta \mu \gg E_C$, the oscillating term in the current saturates and depends only on the charging energy of the Ohmic contact. Thus, the visibility decays with the bias in a power-law manner in this case.  
\end{abstract}

\maketitle

%\section{\label{sec:level1}First-level heading}
% sections are not used for PRL papers

A recent progress in experimental techniques at nano-scale resulted in the emergence of a new field of ``quantum electron optics'', where, as the name suggests,  electrons in one dimensional systems replace photons \cite{eqo}. Typically, one uses for this purpose  QH edge states at integer filling factors, which
play the role of the beams of photons, QPCs, that serve as beam splitters, and Ohmic contacts  to inject equilibrium electrons. Despite many analogies with quantum optics, there is one important difference: quasi-one dimensional electrons strongly interact, and often interactions cannot be considered perturbatively. Not only this requires an application of advanced theoretical methods, such as the bosonization technique \cite{Gogolin,Thierry}, but also leads to a number of new interesting physical effects, most prominently, to the lobe-structure in the visibility of AB oscillations in electronic MZ interferometers \cite{lobes}, the heat Coulomb blockade effect \cite{FP1,Slobodeniuk}, and the saturation of the quantum coherence at high energies \cite{saturation1,saturation2}.

Among various quantum electron optics devices, Ohmic contacts are the most intriguing systems, because in the context of the  QH physics they present an example of strong coupling between completely different states of matter. This is especially the case at fractional fillings, where the strong theoretical effort has already been made \cite{fractional}, while experimentally the physics of Ohmic contacts is far from being fully understood. 
At integer filling factors, one often refers to earlier works of M.\ B\"uttiker, who proposed to consider Ohmic contacts as analogues of a black body in quantum optics, i.e., a reservoir of equilibrium electrons. According to the voltage probe \cite{voltage} and dephasing probe \cite{dephasing} models, edge electrons entering an Ohmic contact are fully equilibrated, or  loose completely their phase coherence. These models have been widely used in the literature \cite{Blanter}.

\begin{figure}
\includegraphics[scale=0.28]{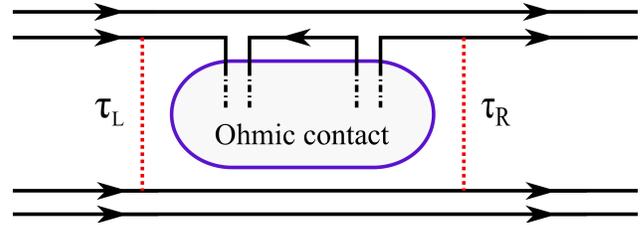}
\caption{\label{fig:one} An Ohmic contact is a piece of metal, which is placed in close vicinity to the 2DEG. It absorbs incoming electron states, formed at the edge of the 2DEG in the QH effect regime (shown by thick black lines) and turns them to neutral electron-hole excitations (dotted lines) and charge fluctuations. Then, it  emmits equilibrium neutral excitations as well as the charge current into the outgoing edge states. Here, the case where only one channel at filling factor 2 is transmitted to the Ohmic contact is shown. The Ohmic contact is embedded into one of the arms of a MZ interferometer, which is formed between two QPCs, shown by red dotted lines, where edge states are weakly mixed with amplitudes $\tau_L$ and $\tau_R$. The bias $\Delta \mu$ applied to the upper chiral channel causes the tunneling current $\langle I\rangle$ to the lower channel. The differential conductance associated with this current $G=\partial_{\Delta \mu} \langle I \rangle $ oscillates as a function of the AB phase $\phi_{\text{AB}}$. The length of lower arm, $L$, is assumed to be small, namely, $\Delta \mu L/v_{F} \ll 1$ and $T L/v_{F} \ll 1$, where $T$ is the bath temperature.}
\end{figure}

Such approach has some grounds in the case of chiral edge fermions at integer filling factors, where local correlation functions coinside with those for free fermions \cite{Levkivskyi}. However, as has been pointed out in the Ref.\ \cite{Slobodeniuk}, in contrast to photons, electrons carry electrical charge, which has to be taken into account, if one considers an Ohmic contact from somewhat broader perspective as a  metallic island strongly coupled to edge states \cite{FP2,FPMatveev}. Indeed, 
even if the level spacing in such systems is negligible, the charging energy may compare to the base temperature of the experiment and other characteristic energies. As a result, such an Ohmic contact cannot fully equilibrate edge electrons \cite{Slobodeniuk}. A related phenomenon of the Coulomb blockade of the heat flux has recently been observed in the experiment \cite{FP1}. 

%\begin{figure}
%\includegraphics[scale=0.22]{figure1.pdf}
%\includegraphics[scale=0.28]{figure1.eps}
%\caption{\label{fig:one} An Ohmic contact is a piece of metal, which is placed in close vicinity to the 2DEG. It absorbs %incoming electron states, formed at the edge of the 2DEG in the QH effect regime (shown by thick black lines) and turns %them to neutral electron-hole excitations (dotted lines) and charge fluctuations. Then, it  emmits equilibrium neutral %excitations as well as the charge current into the outgoing edge states. Here, the case where only one channel at filling %factor 2 is transmitted to the Ohmic contact is shown. The Ohmic contact is embedded into one of the arms of a MZ %interferometer, which is formed between two QPCs, shown by red dotted lines, where edge states are weakly mixed with %amplitudes $\tau_L$ and $\tau_R$. The bias $\Delta \mu$ applied to the upper chiral channel causes the tunneling current %$\langle I\rangle$ to the lower channel. The differential conductance associated with this current $G=\partial_{\Delta\mu} \langle I \rangle $ oscillates as a function of the AB phase $\phi_{\text{AB}}$. The length of lower arm, $L$, is %assumed to be small, namely, $\Delta \mu L/v_{F} \ll 1$ and $T L/v_{F} \ll 1$, where $T$ is the bath temperature.}
%\end{figure}

In this Letter, we consider the dephasing probe model of an Ohmic contact and show that it fails in a particular case, where a metallic island with a finite charging energy is coupled to an arm of a MZ interferometer via  a QPC that transmits exactly one channel, as shown in Fig.\ \ref{fig:one}. This is a rather strong statement, because it implies that electrons that enter and exit a metallic island are not statistically and quantum mechanically independent, despite the fact that the level spacing inside the island vanishes, and it can be considered a reservoir of neutral modes. This is because according to the effective theory of QH edge states \cite{Wen} the phase differece of incoming and outgoing edge electrons is fully determined by the charge of the metallic island, and thus at energies lower than its charging energy there is no room for phase fluctuations. 
This statement is particularly interesting  in the context of the existing hydrodinamic theory \cite{Aleiner},  suggesting that each edge electron carries infinte number of neutral modes. Such a scenario has recently been investigated with the help of the heat flux measurenets with somewhat inconclusive results \cite{FP3}.
Thus, our proposal could also be considered as an ultimate test of the effective theory of QH edge states.

%\begin{figure}
%\includegraphics[scale=0.22]{figure1.pdf}
%\includegraphics[scale=0.28]{figure1.eps}
%\caption{\label{fig:one} An Ohmic contact is a piece of metal, which is placed in close vicinity to the 2DEG. It absorbs %incoming electron states, formed at the edge of the 2DEG in the QH effect regime (shown by thick black lines) and turns %them to neutral electron-hole excitations (dotted lines) and charge fluctuations. Then, it  emmits equilibrium neutral %excitations as well as the charge current into the outgoing edge states. Here, the case where only one channel at %filling factor 2 is transmitted to the Ohmic contact is shown. }
%\end{figure}

\begin{figure}
\includegraphics[scale=0.28]{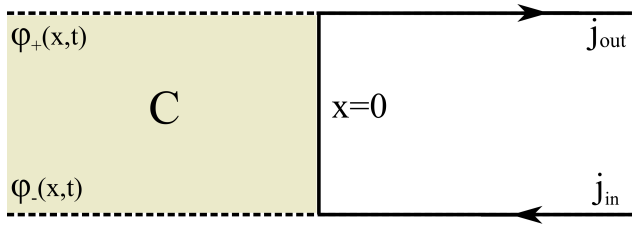}
\caption{\label{fig:two} An equivalent representation of the Ohmic contact at filling factor $\nu =1$. Edge states are described by two bosonic fields $\phi_{+}(x)$ and $\phi_{-}(x)$ of opposite chiralities. The region inside the Ohmic contact, where the capacitive interaction is assumed, is shown by the light yellow color.}
\end{figure}

%===============================================================================================================
\textit{Model of Ohmic contact}. 
We consider an Ohmic contact as a piece of disordered metal of the finite geometrical capacitance $C$ strongly coupled to a QH edge  \cite{Footnote}. 
 We assume a capacitive interaction of electrons inside the Ohmic contact. The level spacing of neutral modes in it is negligible, while its charging energy $E_C=e^2/2C$ is finite. To take into account this fact, we follow the steps outlined in Ref.\ \cite{Slobodeniuk} and model neutral modes by elongating the electron channel inside  the Ohmic contact to infinity, spliting it in two uncorrelated channels, and introducing the regularisation parameter $\varepsilon$ in the Eq.\ (\ref{Operator of total charge accumulated at the Ohmic contact}).  This is schematically shown
in Fig.\ \ref{fig:two}. Throughout the paper, we set $e=\hbar=k_B=1$. 

We use the low-energy effective theory to describe the  QH edge states \cite{Wen}. According to this theory, collective fluctuations of the charge densities  in the electron channels  $\rho_{\sigma}=(1/2\pi)\partial_x \phi_{\sigma}$ are expressed in terms of the bosonic fields
$\phi_{\sigma}(x,t)$, 
where the index $\sigma=-,+$ stands for incoming and outgoing channels, respectively.  The bosonic fields satisfy standard canonical commutation relation:
\begin{equation}
\label{Canonical commutation relation for bosonic fields}
[\partial_x \phi_{\sigma}(x,t),\phi_{\sigma^{\prime}}(y,t)]=2\pi i\sigma \delta_{\sigma \sigma^{\prime}}\delta(x-y).
\end{equation}  
The Hamiltonian of the system consisting of edge states strongly coupled to the Ohmic contact includes two terms
\begin{equation}
\label{Hamiltonian of edge states connected to Ohmic contact}
H=\frac{v_F}{4\pi}\sum_{\sigma} \int\limits_{-\infty}^\infty dx(\partial_x \phi_{\sigma})^2+\frac{Q^2}{2C},
\end{equation}
where
\begin{equation}
\label{Operator of total charge accumulated at the Ohmic contact}
Q=\int_{-\infty}^0 dx e^{\varepsilon x/v_F}[\rho_{+}(x)+\rho_{-}(x)]
\end{equation}
is an operator of the total charge accumulated at the Ohmic contact, and $\varepsilon$
is the regularization parameter. The first term in Eq.\ (\ref{Hamiltonian of edge states connected to Ohmic contact})
is the kinetic energy part of the Hamiltonian, which describes the dynamics of incoming and outgoing edge channels. The second term is the charging 
energy of the Ohmic contact of a finite size.

%\begin{figure}
%\includegraphics[scale=0.32]{figure2.pdf}
%\includegraphics[scale=0.28]{figuretwo.eps}
%\caption{\label{fig:two} An equivalent representation of the Ohmic contact at filling factor $\nu =1$. Edge states are %described by two bosonic fields $\phi_{+}(x)$ and $\phi_{-}(x)$ of opposite chiralities. The region inside the Ohmic %contact, where the capacitive interaction is assumed, is shown by the light yellow color.}
%\end{figure}

Using commutation relations (\ref{Canonical commutation relation for bosonic fields}) and the Hamiltonian (\ref{Hamiltonian of edge states connected to Ohmic contact}), we write the equations of motion for the bosonic fields $\phi_{\sigma}(x,t)$:
\begin{equation}
\label{Equation of motion for bosonic fields}
\sigma \partial_t \phi_{\sigma}+v_F\partial_x\phi_{\sigma}=-\frac{Q(t)e^{\varepsilon x/v_F}}{C}\theta(-x).
\end{equation}
These equations have to be complemented with the following boundary 
conditions:
\begin{equation}
\label{Boundary conditions for bosonic fields}
\begin{split}
& \partial_t \phi_{+}(-\infty,t)=-2\pi j_{\text{s}}(t),\\
& \partial_t \phi_{-}(0,t)=2\pi j_{\text{in}}(t),
\end{split}
\end{equation}
where $j_{\text{in}}(t)$ is the current flowing into the Ohmic contact, while $j_s(t)$ (source current) describes
equilibrium fluctuations of the neutral mode with the temperature $T$. Solving equations (\ref{Equation of motion for bosonic fields}) with the boundary conditions (\ref{Boundary conditions for bosonic fields}), one relates the outgoing current $j_{\text{out}}=-\partial_t \phi_{+}(0,t)/2\pi$ to the incoming current $j_{\text{int}}(t)$, as shown in Fig.~\ref{fig:two}:
\begin{equation}
\label{Outgoing current}
j_{\text{out}}(\omega)=\frac{i\omega R_{q}C }{i\omega R_{q}C-1}j_{s}(\omega)-\frac{1}{i\omega R_{q}C-1}j_{\text{in}}(\omega),
\end{equation}
where $R_q=2\pi \hbar/e^2$ is a quantum of resistance (restoring natural unites).

The statistics of current fluctuations  $\delta j_{\alpha}(\omega)\equiv j_{\alpha}(\omega)-\langle j_{\alpha}(\omega)\rangle$, where $\alpha=\text{in},\text{s}$, is characterized by the equilibrium density function $S(\omega)$~\cite{Lifshitz}
\begin{equation}
S(\omega)=\frac{\omega/R_{q}}{1-e^{-\omega/T}}
\end{equation}
defined via the relation 
\begin{equation}
\label{eqcorr}
\langle \delta j_{\alpha}(\omega)\delta j_{\beta}(\omega^{\prime})\rangle=2\pi \delta_{\alpha \beta}\delta(\omega+\omega^{\prime})S(\omega).
\end{equation}
Eqs.\ (\ref{Boundary conditions for bosonic fields}-\ref{eqcorr}) can now be used to calculate two-point correlation functions of the fields $\phi_\sigma(x,t)$. 

%============================================================================================================
\textit{Electronic Mach-Zehnder interferometer}.
%\begin{figure}
%\includegraphics[scale=0.25]{figure3.pdf}
%\includegraphics[scale=0.28]{figure3.eps}
%\caption{\label{fig:three} Schematic representation of the electronic MZI at filling factor 1 coupled to an Ohmic %contact is shown. Two edge states propagating from left to right are mixed by two quantum point contacts and form the AB %loop. The bias $\Delta \mu$ applied to the upper chiral channel causes the tunneling current $\langle I\rangle$ to the %lower channel. The differential conductance associated with this current $G=\partial_{\Delta \mu} \langle I \rangle $ %oscillates as a function of the AB phase $\phi_{\text{AB}}$. The length of lower arm, $L$, is assumed to be small, %namely, $\Delta \mu L/v_{F} \ll 1$ and $T L/v_{F} \ll 1$, where $T$ is the bath temperature. }
%\end{figure}
Schematic representation of an electronic MZI attached to an  Ohmic contact  is shown in Fig.~\ref{fig:one}. Two point contacts located at positions $x_l$, where $l=L,R$, mix the edge states and allow interference between them. This can be described by the tunneling Hamiltonian with the vertex operators at $x_l$~\cite{Levkivskyi}
\begin{equation}
\label{tunneling}
\begin{split}
& H_T=A_L+A_R+ \text{H.c.}, \\
& A_l \propto \tau_l \exp[-i\phi_{u}(x_l)+i\phi_{d}(x_l)],
\end{split}
\end{equation}
where $\tau_l$ are the tunneling coupling amplitudes, and $\phi_i(x_l)$, $i=\text{u},\text{d}$, are the bosonic fields  at the upper and lower channel of the MZ interferometer, respectively. The AB phase is included in the tunneling amplitudes  via the relation $\tau^{\ast}_R\tau_L=|\tau_R||\tau_L|e^{i\phi_{\text{AB}}}$.

 We investigate interference effects in the electron tunneling current.  It is defined as a rate of change of the electron number $N_{\text{d}}$ in the lower arm, $I=i[H,N_{\text{d}}]$. To the leading order in tunneling amplitudes, its average value is given by the Kubo linear response formula
\begin{equation} 
\label{Kubo formula for current}
\langle I\rangle=\int\limits_{-\infty}^\infty dt \sum_{l,l^{\prime}}\langle [A^{\dagger}_{l}(t),A_{l^{\prime}}(0)]\rangle, 
\end{equation}  
where the average is taken with respect to the equilibrium state of the system biased by the potential difference $\Delta \mu$. As one can easily see, the total current consists of three terms: $\langle I\rangle=I_{\text{LL}}+I_{\text{RR}}+2\text{Re}(I_\text{RL})$,  where the third term contains the AB phase. The degree of the phase coherence is characterized by the visibility of AB oscillations
\begin{equation}
\label{Visibility}
V_{\text{AB}}=\frac{G_{\text{max}}-G_{\text{min}}}{G_{\text{max}}+G_{\text{min}}},
\end{equation}
where $G=\partial_{\Delta \mu}\langle I \rangle$ is the differential conductance associated with tunneling current $\langle I\rangle$. In the rest of the paper we investigate the dependence of the visibility on the temperature $T$ and applied bias $\Delta \mu$. The details of the calculations are presented in the supplementary material. Here, we mention only that in order to evaluate the average current (\ref{Kubo formula for current}), we use Eqs.\ (\ref{Boundary conditions for bosonic fields}-\ref{tunneling}) and the Gaussian character of the theory. 

%========================================================================================================================
\textit{Direct current and conductance}. 
In our model the interaction is present only in the Ohmic contact located between points $x_L$ and $x_R$. The important consequence of this fact is that the interaction cannot affect the direct contribution $I_{\text{dir}}=I_{\text{LL}}+I_{\text{RR}}$, which also from the unitarity of scattering relation (\ref{Outgoing current}). Therefore, we readily obtain the direct part of total current
\begin{equation}
\label{Incoherent current. Linear regime}
I_{\text{dir}}=\frac{|\tau_L|^2+|\tau_R|^2}{2\pi v^2_{F}}\Delta \mu.
\end{equation}
Thus, the direct conductance $G_{\text{dir}}=\partial_{\Delta \mu}I_{\text{dir}}$,
\begin{equation}
\label{Incoherent conductance. Linear regime}
G_{\text{dir}}=\frac{|\tau_L|^2+|\tau_R|^2}{2\pi v^2_{F}}
\end{equation}
shows conventional ohmic behavior, i.e., it is independent of the temperature $T$ and bias $\Delta\mu$. 

%========================================================================================================================
\textit{Visibility of AB oscillations. Low-bias Ohmic regime}.  We  consider the oscillating part of the conductance, $G_{\text{osc}}\equiv\partial_{\Delta\mu}[2\text{Re}(I_\text{RL})]$, focus on the regime of low bias, and take the limit $\Delta \mu \to 0$.  At low temperatures, $T\ll E_C$, the behavior is the same as for non-interacting fermions~\cite{Supplementary material}:
\begin{equation}
\label{Coherent conductance. Linear regime. Low temperature case}
G_{\text{osc}}=\frac{|\tau_L||\tau_R|}{\pi v^2_{F}}\cos(\phi_{AB}),
\end{equation}
and according to the  Eq.~(\ref{Visibility}) the visibility acquires the following form:
\begin{equation}
\label{Visibility.Linear regime. Low temperature case}
V\equiv V_0=\frac{2|\tau_L||\tau_R|}{|\tau_L|^2+|\tau_R|^2}.
\end{equation}
Thus, at low temperatures, 
$T \ll E_C$, thermal fluctuations are not able to suppress the quantum coherence of edge states despite the fact that they are strongly coupled to an Ohmic contact.

In the opposite limit of high temperatures, $T \gg E_C$, we obtain the following result for the oscillating part of the conductance~\cite{Supplementary material}
\begin{equation}
\label{Coherent conductance. Linear regime. High temperature case}
G_{\text{osc}}=\frac{|\tau_L||\tau_R|}{v^2_{F}}\sqrt{\frac{\pi T}{E_C}}e^{-\pi^2 T/E_C}\cos(\phi_{\text{AB}}).
\end{equation}
 Next, substituting Eqs.\ (\ref{Incoherent conductance. Linear regime}) and (\ref{Coherent conductance. Linear regime. High temperature case}) into the Eq.\ (\ref{Visibility}), we obtain 
\begin{equation}
\label{Visibility. Linear regime. High temperature case}
V/V_0=\pi\sqrt{\frac{\pi T}{E_C}}e^{-\pi^2 T/E_C}.
\end{equation}
\begin{figure}
\includegraphics[scale=0.35]{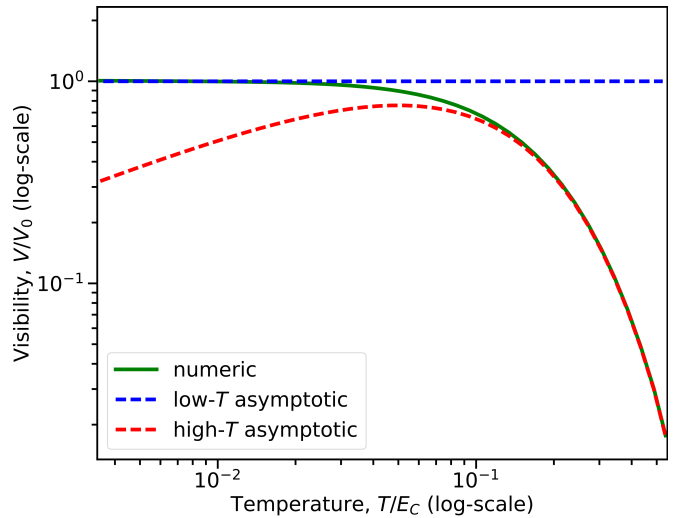}
\caption{\label{fig:four} The normalized visibility $V/V_0$ is plotted versus the dimensionless temperature $T/E_C$ in log-log scale.}
\end{figure}
The dependence of $V/V_0$ on temperature is given in Fig.~\ref{fig:four}.

%====================================================================================================================
\textit{Visibility of AB oscillations. Nonlinear regime}. In this section we focus on the nonlinear regime, namely $\Delta \mu$ is arbitrary, and $T \to 0$. In the case of small bias,  $\Delta \mu \ll E_C$, the free-fermionic behavior is restored~\cite{Supplementary material}, and the visibility is given by Eq.\ (\ref{Visibility.Linear regime. Low temperature case}). In the case of the large bias, $\Delta \mu \gg E_C$, we obtain the following result \cite{Supplementary material} for the oscillating (coherent) part of the current, $I_{\text{osc}}=2\text{Re}(I_\text{RL})$, including the sub-leading term in the bias:
%\begin{equation}
%\label{Coherent current. Nonlinear regime}
%\begin{split}
%& I_{\text{osc}}=I_1+I_2, \\
%& I_1=\frac{|\tau_L||\tau_R|}{2\pi v^2_{F}}\frac{2e^{\bold{\gamma}}E_C}{\pi}\cos(\phi_{\text{AB}}+\pi/2), \\
%& I_2=\frac{|\tau_L||\tau_R|}{2\pi v^2_{F}}\frac{2e^{\bold{\gamma}}E_C}{\pi}\frac{E_C}{\pi \Delta \mu}\cos(\phi_{\text{AB}})
%\end{split}
%\end{equation}
\begin{equation}
\label{Coherent current. Nonlinear regime}
\begin{split}
& I_{\text{osc}}=\frac{|\tau_L||\tau_R|}{2\pi v^2_{F}}\frac{2e^{\bold{\gamma}}E_C}{\pi}\\
&\times\left[\cos(\phi_{\text{AB}}+\pi/2)+\frac{E_C}{\pi \Delta \mu}\cos(\phi_{\text{AB}})\right],
\end{split}
\end{equation}
where $\bold{\gamma}\approx 0.5772$ is an Euler constant. Interestingly, at biases larger than the charging energy $E_C$ of the Ohmic contact the coherent contribution to the current saturates at values $\propto E_C$. This  implies, that it possibly originates from the elestic tunneling induced by the resonance in the transmission of plasmons in the upper arm of the interferometer. Combining Eqs.\ (\ref{Visibility}), (\ref{Incoherent current. Linear regime}) and (\ref{Coherent current. Nonlinear regime}), one arrives at the following expression for the visibility of AB oscillations 
\begin{equation}
\label{Visibility. Nonlinear regime}
V/V_0=\frac{e^{\bold{\gamma}}E^2_C}{(\pi \Delta \mu)^2}.
\end{equation}
\begin{figure}
\includegraphics[scale=0.35]{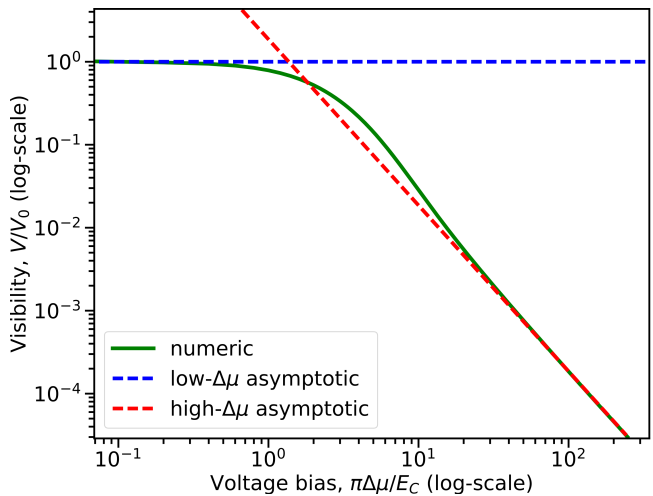}
\caption{\label{fig:five} The normalized visibility $V/V_0$ is plotted versus the dimensionless bias $\pi \Delta \mu /E_C$ in log-log scale.}
\end{figure}
The full dependence of $V/V_0$ on the bias is shown in Fig.~\ref{fig:five}.

Finally, we note that at filling factor $\nu\geq 2$, or more generally, when an Ohmic contact is perfectly coupled to at least two electron channels, the phase coherence is fully suppressed, as shown in the supplementary material~\cite{Supplementary material}. This can easily be explained by the fact, that it does not cost  energy to flip the pseudo-spin related to extra electron channels, because the level spacing of neutral modes in the Ohmic contact is assumed to be zero.
%=======================================================================================================================

To \textit{summarize}, we have studied the dephasing mechanism in the electronic MZ interferometer based on the edge states in a QH system at filling factor $\nu=1$, strongly coupled to an Ohmic contact. Alternatively,  an Ohmic contact may be connected to an interferometer by a QPC, transmitting only one electron channel. We have used a simple model of an Ohmic contact as a reservoir of neutral modes with the finite charging energy $E_C$. It was shown earlier \cite{Slobodeniuk} that such an Ohmic contact is  not always able to  fully equilibrate edge states. Here, we have demonstrated that it is also not always able to fully  suppress the phase coherence of edge electrons. This is because edge electrons carry charge, and at  filling factor $\nu=1$ the phase of an electron is determined by the charge occumulated at an Ohmic contact. At temperatures and voltage biases smaller than the charging energy $E_C$, charge fluctuations, and consequently phase fluctuations are suppressed. On the other hand, at filling factors larger than 1 additional degrees of freedom of the edge electrons are perfectly coupled to neutral modes in the Ohmic contact, which leads to the full suppression of the phase coherence \cite{Supplementary material}.

We acknowledge support from the Swiss NSF.
%======================================================================================================================

%\input acknowledgement.tex   % input acknowledgement


\begin{thebibliography}{99}
\bibitem{Footnote}
Because of strong coupling of the Ohmic contact to edge state there is no sign of Coulomb blockade phenomenon in present
model
\end{thebibliography}
\end{document}

\title{Supplementary material ``Dephasing in Mach-Zehnder interferometer by an Ohmic contact''}

\author{Edvin G. Idrisov}
\affiliation{D\'epartement de Physique Th\'eorique, Universit\'e de Gen\`eve, CH-1211 Gen\`eve 4, Switzerland}

\author{Ivan P. Levkivskyi}
\affiliation{Theoretische Physik, ETH Zurich, CH-8093 Zurich, Switzerland}
\affiliation{Institute of Ecology and Evolution, University of Bern, CH-3012 Bern, Switzerland}
\affiliation{Department of Computational Biology, University of Lausanne, CH-1011 Lausanne, Switzerland}

\author{Eugene V. Sukhorukov}
\affiliation{D\'epartement de Physique Th\'eorique, Universit\'e de Gen\`eve, CH-1211 Gen\`eve 4, Switzerland}
\date{\today}

% The following information is for internal review, please remove them for submission
%\widetext
%\leftline{Version xx as of \today}
%\leftline{Primary authors: Joe E. Physics}
%\leftline{To be submitted to (PRL, PRD-RC, PRD, PLB; choose one.)}
%\leftline{Comment to {\tt d0-run2eb-nnn@fnal.gov} by xxx, yyy}
%\centerline{\em D\O\ INTERNAL DOCUMENT -- NOT FOR PUBLIC DISTRIBUTION}

% the following line is for submission, including submission to the arXiv!!
%\hspace{5.2in} \mbox{Fermilab-Pub-04/xxx-E}

%\title{Template for PRL/PRD Papers}
%\input author_list.tex       % D0 authors (remove the first 3 lines
                             % of this file prior to submission, they
                             % contain a time stamp for the authorlist)
                             % (includes institutions and visitors)
%\date{\today}

%\pacs{}
\maketitle

%\section{\label{sec:level1}First-level heading}
% sections are not used for PRL papers

%============================================================================================
\subsection{Visibility of AB oscillations: Linear bias regime}
As we have mentioned in the main text of the present paper, finding the direct contribution to the total current (conductance) is straightforward, and the result is given by Eq.(12-13). The main task is to calculate the oscillating part of conductance, namely $G_{\text{osc}}=\partial_{\Delta\mu}[2\text{Re}(I_\text{RL})]$. Taking the derivative with respect to the applied bias and setting $\Delta \mu =0$, we get  
\begin{equation}
\label{SCoherent conductance. Linear regime}
G_{\rm osc}=2|\tau_R||\tau_L| C_{RL}\cos(\phi_{AB}),
\end{equation}
where the factor $C_{RL}$ includes effects of an Ohmic contact and  has the following form 
\begin{equation}
\label{CRL integral}
C_{RL}=\int\limits_{-\infty}^\infty dt it \left[K_1(t) - K_2(t)\right]. 
\end{equation}
In our model the upper and lower arms of the MZI do not interact, therefore the correlation functions $K_1(t)$ and $K_2(t)$ split into the product of two single-particle correlators:
\begin{equation}
\label{free fermion correlators}
\begin{split}
& K_1(t)=\frac{1}{a^2}\langle e^{-i\phi_u(x_R=0,t)}e^{i\phi_u(x_L=0,0)}\rangle \langle e^{i\phi_d(x_R=0,t)}e^{-i\phi_d(x_L=0,0)}\rangle, \\
& K_2(t)=\frac{1}{a^2}\langle e^{i\phi_u(x_L=0,0)}e^{-i\phi_u(x_R=0,t)}\rangle \langle e^{-i\phi_d(x_L=0,0)}e^{i\phi_d(x_R=0,t)}\rangle,
\end{split}
\end{equation}
where $a$ is the utraviolet cut-off, and free-fermion correlation functions at different times are given by 
\begin{equation}
\begin{split}
& \frac{1}{a^2}\langle e^{i\phi_d(x_R=0,t)}e^{-i\phi_d(x_L=0,0)}\rangle =\frac{-iT}{2v_F\sinh[\pi T(t-i0)]},\\
& \frac{1}{a^2}\langle e^{-i\phi_d(x_L=0,0)}e^{i\phi_d(x_R=0,t)}\rangle = \frac{iT}{2v_F\sinh[\pi T(t+i0)]}.\\
\end{split}
\end{equation}
Correlation functions that depend on the Ohmic contact energy scales are calculated using the expressions (5-9) of the main text and the Gaussian character of the field fluctuations:
\begin{equation}
\begin{split}
& \frac{1}{a^2}\langle e^{-i\phi_u(x_R=0,t)}e^{i\phi_u(x_L=0,0)}\rangle =\frac{1}{a^2}\exp\left(\int \frac{d\omega \omega}{\omega^2+\eta^2}\frac{t(\omega)e^{-i\omega t}-1}{1-e^{-\omega/T}}\right),\\
& \frac{1}{a^2}\langle e^{i\phi_u(x_L=0,0)}e^{-i\phi_u(x_R=0,t)}\rangle = \frac{1}{a^2}\exp\left(\int \frac{d\omega \omega}{\omega^2+\eta^2}\frac{t^{*}(\omega)e^{i\omega t}-1}{1-e^{-\omega/T}}\right),
\end{split}
\end{equation}
where $\eta \to +0$, and the transmission coefficient for the edge magnetoplasmons has the form $t(\omega)=1/(1-i\omega R_q C)$.

 Next, we consider only the first integral, because the second one differs only by the complex conjugation. We rewrite the integral in the form suitable for the integration
\begin{align}
\int \frac{d\omega \omega}{\omega^2+\eta^2}\frac{t(\omega)e^{-i\omega t}-1}{1-e^{-\omega/T}}=a(T)+\beta(t)+F(t),
\end{align}
where 
\begin{equation}
a(T)=iR_qC\int \frac{d\omega \omega^2}{\omega^2+\eta^2}\frac{1}{1-i\omega R_qC}\frac{1}{1-e^{-\omega/T}},
\end{equation}
\begin{equation}
\label{beta integral}
\beta(t)=\int \frac{d\omega \omega}{\omega^2+\eta^2}\frac{1}{1+(\omega R_qC)^2}\frac{e^{-i\omega t}-1}{1-e^{-\omega/T}},
\end{equation}
\begin{equation}
\label{F integral}
F(t)=\int \frac{d\omega \omega}{\omega^2+\eta^2}\frac{i\omega R_qC}{1+(\omega R_qC)^2}\frac{e^{-i\omega t}-1}{1-e^{-\omega/T}}.
\end{equation}
Here the first integral is time independent and can be calculated exactly: $a(T)=i \pi /2 + \alpha(T)$, where
\begin{equation}
\alpha(T)=\bold{\gamma}+\log \left(\frac{a}{2\pi v_F R_qC} \right)+\pi TR_qC+\log(2\pi TR_qC)+\Psi(1/2\pi TR_qC).
\end{equation}
Here,
$\bold{\gamma}\approx 0.5772$ is the Euler constant, $\Psi(x)= d \log(\Gamma(x))/dx$ is the digamma function of real variable $x$ and $\Gamma(x)$ is the gamma function.
%=================================================================================================================
\subsubsection{High temperatures, $T\gg E_C$}
At high temperatures integrals in Eqs.~(\ref{beta integral}) and (\ref{F integral}) can be calculated by expanding the bosonic distribution function, $1/(1-\exp(-\omega/T))\approx T/\omega + 1/2$. We obtain
\begin{equation}
\beta(t)=\pi TR_qC\left(1-e^{-|t/R_qC|}-|t/R_qC|\right)-\frac{i\pi}{2} \left(1-e^{-|t/R_qC|}\right)\text{sign}(t),
\end{equation}
\begin{equation}
F(t)=\pi TR_qC\left(1-e^{-|t/R_qC|}\right)\text{sign}(t)-\frac{i\pi}{2} \left(1-e^{-|t/R_qC|}\right).
\end{equation}
By substituting Eqs.\ (\ref{free fermion correlators}-\ref{F integral}) into Eq.~(\ref{CRL integral}) for $C_{RL}$, we get 
\begin{equation}
\label{CRL integral2}
C_{RL}=\frac{T^2e^{\alpha(T)}}{2v^2_F}\int\limits_{-\infty}^\infty dt it \left(\frac{e^{\beta(t)+F(t)}}{\sinh[\pi T(t-i0)]}-\frac{e^{\beta^{\star}(t)+F^{\star}(t)}}{\sinh[\pi T(t+i0)]}\right),
\end{equation}
where we used the limit $\alpha(t)= a/2\pi v_F R_qC -\pi TR_qC +\log(2\pi TR_qC)$ at $TR_qC \gg 1$.

 Next, we split this integral into negative and positive time contributions. For  $t<0$, using the expressions (\ref{beta integral}) and (\ref{F integral}), one can show that $\beta(t)+F(t)=\pi Tt$. Therefor, the negative time contribution vanishes
\begin{equation}
\int_{-\infty}^0 dx ix e^{x}\left[\frac{1}{\sinh[x-i0]}-\frac{1}{\sinh[x+i0]}\right]=0,
\end{equation}
where $x=\pi Tt$. 
For  $t>0$, the main contribution to the time  integral comes from times $t/R_qC \ll 1$ and $\pi Tt \gg 1$. After expanding in $t/R_qC$ one gets $\beta(t)+F(t)\approx -\pi Tt^2/R_qC -i\pi t/R_qC + \pi Tt$, and the Eq.\ (\ref{CRL integral2}) simplifies:
\begin{equation}
C_{RL}=\frac{\pi T^2e^{-\pi TR_qC}}{v^2_F R_qC}\int_0^{\infty} dt t^2 e^{-\pi Tt^2/R_qC}\frac{e^{\pi Tt}}{\sinh(\pi Tt)}.
\end{equation} 
We take the limit $\exp(\pi Tt)/\sinh(\pi Tt)= 2$ at $\pi Tt \gg 1$ and use again the dimensionless variable $x=\pi Tt$ to write:
\begin{equation}
C_{RL}=\frac{2}{\pi v^2_F}\frac{e^{-\pi TR_qC}}{\pi TR_qC}\int_0^{\infty} dx x^2 e^{-x^2/\pi TR_qC}.
\end{equation}
Finally, evaluating the integral and substituting the result into $G_{\text{osc}}$ given by the Eq.\ (\ref{SCoherent conductance. Linear regime}), we obtain   
\begin{equation}
G_{osc}=\frac{|\tau_L||\tau_R|}{2\pi v^2_F}2\pi \sqrt{\frac{\pi T}{E_C}}e^{-\pi^2 T/E_C}\cos(\phi_{AB}). 
\end{equation}

The total conductance, $G=G_{dir}+G_{osc}$, at high temperatures $T\gg E_C$ reads
\begin{equation}
G=\frac{|\tau_L|^2+|\tau_R|^2}{2\pi v^2_F}+\frac{|\tau_L||\tau_R|}{2\pi v^2_F}2\pi \sqrt{\frac{\pi T}{E_C}}e^{-\pi^2 T/E_C}\cos(\phi_{AB}) .
\end{equation}
Thus, we arrive at the Eq.\ (17) for visibility of AB oscillations in main text
%\begin{equation}
%V=\frac{2|\tau_L||\tau_R|}{|\tau_L|^2+|\tau_R|^2} \pi \sqrt{\frac{\pi T}{E_C}}e^{-\pi^2 T/E_C} 
%\end{equation}
%\subsubsection{Low temperature regime}
%The calculations are similar to the low bias case in the non-linear regime (see below). The free fermion picture is restored.
 The exact curve for visibility, shown in Fig.\ (4) of main text, is obtained by evaluating time integrals numerically.

\subsection{Visibility of AB oscillations: Zero temperature limit} 
In the case of zero temperature we get the following expression for the oscillating  term in the total current
\begin{equation}
\label{SCoherent current}
I_{osc}=2{\rm Re}(I_{\rm RL})=\frac{1}{2\pi^2v^2_F}{\rm Re}(\tau_L\tau^{\ast}_R \tilde{I}),
\end{equation}
where 
\begin{equation}
\label{Itilda integral}
\tilde{I}=\int\limits_{-\infty}^\infty dt e^{i \Delta \mu t}\left[\frac{e^{M(t)}}{(t-i0)^2}-\frac{e^{M^{\ast}(t)}}{(t+i0)^2}\right],
\end{equation}
and
\begin{equation}
M(t)=\int_0^{\infty} dx \frac{i}{1-ix}e^{-ixt/R_qC}.
\end{equation} 
The $\tilde{I}$ can be rewritten to run over positive times, only:
\begin{equation}
\tilde{I}=\int_0^{\infty} dt e^{i\Delta \mu t}\left[\frac{e^{M(t)}}{(t-i0)^2}-\frac{e^{M^{\ast}(t)}}{(t+i0)^2}\right]+
\int_0^{\infty} dt e^{-i\Delta \mu t}\left[\frac{e^{M(-t)}}{(t+i0)^2}-\frac{e^{M^{\ast}(-t)}}{(t-i0)^2}\right] ,
\end{equation}
where $M(t)=i\pi\exp(-t/R_qC)+\exp(-t/R_qC)Ei(t/R_qC)$ and $M(-t)=\exp(t/R_qC)Ei(-t/R_qC)$. Here, we introduced the exponential integral function $Ei(x)=-\int_{-x}^{\infty} dy\exp(-y)/y $.

\subsubsection{Low bias regime}
In this limit $\Delta \mu R_qC \ll 1$ the main contribution to integral in the Eq.\ (\ref{Itilda integral}) comes from large times $t/R_qC \gg 1$, therefore we can use $e^{M(t)}=e^{M(-t)} \sim 1$. This results in the following simple expression for $\tilde{I}$:
\begin{equation}
\tilde{I}=\int_{-\infty}^{\infty} dt e^{i\Delta \mu t}\left[\frac{1}{(t-i0)^2}-\frac{1}{(t+i0)^2}\right]=-2\pi \Delta \mu.
\end{equation}
Consequently, the oscillating part of the current acquires the free-fermionic form
\begin{equation}
I_{osc}=\frac{|\tau_L||\tau_R|}{\pi v^2_F}\Delta \mu,
\end{equation}
thus the visibility of AB oscilations is given by the expression (15) in the main text. 
 
\subsubsection{Large bias regime}
In the limit $\Delta \mu R_qC \gg 1$ the main contribution to the integral (\ref{Itilda integral}) comes from small times $t/R_qC \ll 1$, so we can use $\exp[M(t)]\approx -\frac{e^{\bold{\gamma}}t}{R_qC}(1-i\pi t/R_qC)$ and $\exp[M(-t)]\approx e^{\bold{\gamma}}t/R_qC$, where $\bold{\gamma} \approx 0.5772$ is the Euler constant. As a result, we arrive at  the following expression for $\tilde{I}$
\begin{equation}
\tilde{I}=-\frac{2e^{\bold{\gamma}}}{R_qC}\int_0^{\infty} dx \cos(\Delta \mu x)\left[\frac{1}{x-i0}-\frac{1}{x+i0}\right]+\frac{i\pi e^{\bold{\gamma}}}{\Delta \mu (R_qC)^2}\int_0^{\infty}dx e^{ix -\xi x}\left[\frac{x^2}{(x-i0)^2}+\frac{x^2}{(x+i0)^2}\right],
\end{equation}
where we introduced the dimensionless variable $x=t/R_qC$ and regularized the integral with $\xi \to +0$. Evaluating the integral, we obtain
\begin{equation}
\tilde{I}=-\frac{2\pi  e^{\bold{\gamma}}}{R_qC}\left[i+\frac{1}{\Delta \mu R_qC}\right]. 
\end{equation}
Substituting this result into the Eq.\ (\ref{SCoherent current}) for $I_{osc}$, we get
\begin{equation}
I_{\text{osc}}=\frac{|\tau_L||\tau_R|}{2\pi v^2_{F}}\frac{2e^{\bold{\gamma}}E_C}{\pi}\left[\cos(\phi_{\text{AB}}+\pi/2)+\frac{E_C}{\pi \Delta \mu}\cos(\phi_{AB})\right] .
\end{equation}
After taking the derivative with respect to the bias $\Delta \mu$, substituting the result into the Eq.\ (11) in the main text, we obtain the result (19) of the main text. The exact curve, obtained numerically, is shown in the Fig.\ (5). 

\subsection{Suppression of phase coherence at filling factors $\nu\geq 2$}
At arbitrary filling factor, the equations of motion have the following form
\begin{equation}
\begin{split}
& \dot{Q}(t) = \sum^{\nu}_{\alpha=1} \left[j_{in,\alpha}(t) - j_{out,\alpha}(t)\right],\\
& j_{out, \alpha}(t)=Q(t)/R_q C + j_{s,\alpha}(t), \quad \alpha =1,2,...,\nu.
\end{split}
\end{equation}
Solving this system of equations in frequency representation one relates the outgoing current $j_{out,1}(\omega)$, in channel one to incoming currents $j_{in,\alpha}(\omega)$ and $j_{s,\alpha}(\omega)$
\begin{equation}
\label{out1}
j_{out,1}(\omega)=j_{s,1}(\omega) + \frac{1}{\nu-i\omega R_q C}\sum^{\nu}_{\alpha =1} \left[j_{in,\alpha}(\omega)-j_{s,\alpha}(\omega)\right].
\end{equation}
The statistics of current fluctuations is given by the following expression $\langle \delta j_{\alpha, \beta}(\omega) \delta j_{\gamma, \xi}(\omega^{\prime})\rangle = 2\pi \delta_{\alpha \beta} \delta_{\gamma \xi}\delta (\omega+\omega^{\prime})S(\omega)$, where $\alpha, \gamma =in,s$ and $\beta, \xi =1,2,...,\nu$. Here, the equilibrium density function is given by $S(\omega)=\omega G_q/(1-e^{-\omega/T})$, where $G_q=1/R_q=e^2/h$ is a quantum conductance.

Next, following the steps outlined in the main part of the paper, we find that the total current consists of the direct and oscillating parts: $I=I_{dir}+I_{osc}$. The direct part of the total current does not change, i.e.,   it has a conventional Ohmic behavior
\begin{equation}
I_{dir} = \frac{|\tau_L|^2+|\tau_R|^2}{2\pi v^2_F} \Delta \mu.
\end{equation}
On the other hand, the oscillating part of current, $I_{osc}=2Re[I_{RL}]$, acquires the following form
\begin{equation}
\label{Irl}
I_{RL} = \tau^{\ast}_R \tau_L \int dt e^{i\Delta \mu t}\left[K^2_0(t)\exp[M(t)]-(K^{\ast}_0(t))^2\exp[M^{\ast}(t)]\right].
\end{equation}
Here,  $K_0(t)=-iT/(2v_F\sinh[\pi T(t-i\gamma)])$ is the two point free fermion correlation function, and 
\begin{equation}
\label{M(t)}
M(t)=\int \frac{d\omega \omega}{\omega^2+\eta^2}\frac{[t(\omega)-1]e^{-i\omega t}}{1-e^{-\omega/T}},
\end{equation}
where $t(\omega)=1/(\nu-i\omega R_q C)$ is the scattering coefficient that stands in front of the incoming current, $j_{in,1}(\omega)$ in Eq.\ (\ref{out1}).

We note, that the integral in (\ref{M(t)})  is well defined for $\nu=1$, but in the case of $\nu\geq 2$ it has a divergence at zero frequencies and at arbitrary capacitance $C$ and temperature $T$. Thus,  $\exp[M(t)] \to 0$ at filling factors $\nu \geq 2$, which leads to a full suppression of the phase coherence according to Eq.\ (\ref{Irl}).